\documentclass{ifacconf}  
\usepackage{natbib}
\usepackage{epsfig} 
\usepackage{amsmath,amssymb,amsfonts} 
\usepackage{subfigure}
\usepackage{float, color}
\usepackage{enumitem}
\newcommand{\seb}[1]{%
{\leavevmode\color{black}#1}%
}

\usepackage{theorem}
\theoremstyle{plain} \theorembodyfont{\upshape}
\newtheorem{theorem}{Theorem}
\newtheorem{assm}{Assumption}
\newtheorem{remark}{Remark}
\newtheorem{proposition}{Proposition}

\def\qed{\hfill $\Box$}
\DeclareMathOperator{\diag}{diag}
\usepackage{natbib}  

\allowdisplaybreaks[1]
\usepackage{mathtools}
\begin{document}

\thispagestyle{empty}
\pagestyle{plain}

\newcommand{\cu}[1]{%
{\leavevmode\color{black}#1}%
}
\newcommand{\cub}[1]{%
{\leavevmode\color{black}#1}%
}

\allowdisplaybreaks

\begin{frontmatter}

\title{
Multi-Competitive Virus Spread over a Time-Varying Networked SIS Model with an Infrastructure Network}


\author[First]{Sebin~Gracy}
\author[Second]{Yuan~Wang}
\author[Third]{Philip E.~Par\'e}
\author[First]{C\'esar A Uribe}

\address[First]{
Department of Electrical and Computer Engineering, Rice University, Houston, TX, USA. sebin.gracy@rice.edu, cauribe@rice.edu}
 \address[Second]{ Department of Robotics, Hunan University, Changsha, China. yuanw@hnu.edu.cn}
\address[Third]{Elmore School of Electrical and Computer Engineering, Purdue University, West Lafayette, IN, USA. philpare@purdue.edu}


\begin{abstract}
We study the spread  of multi-competitive viruses over a (possibly) time-varying network of individuals accounting for the presence of shared infrastructure networks that further enables transmission of the virus. We establish a sufficient condition for exponentially fast eradication of a virus for: 1) 
time-invariant graphs, 2) time-varying graphs with symmetric interactions between individuals and homogeneous 
virus spread across the 
network (same healing and infection rate for all individuals), and 3) directed and slowly varying graphs with  heterogeneous virus spread (not necessarily same
healing and infection rates for all individuals) across the 
network. Numerical examples illustrate our theoretical results and indicate that, for the time-varying case, violation of the aforementioned sufficient conditions could lead to the persistence of a virus.        
\end{abstract}


\end{frontmatter}

\section{Introduction}\label{sec:intro}

 The social and economic impacts of epidemics and their higher-order effects are enormous~\citep{johnson2002updating}. Prominent cases of epidemics include the Spanish flu $1918$--$1920$ and the Asian flu in the $1950$s~\citep{jackson2009history}. Although modeling, analysis, and control of the spread of (biological) viruses have been studied for several decades ~\citep{van2008virus,bloom2018epidemics,hethcote2000mathematics,nowzari2016analysis}, the current COVID-19 crisis has sparked increasing interest recently~\citep{giordano2020modelling}. 
 Existing research tries to understand what causes a disease to spread, how the spread can be mitigated or eradicated, and how to estimate infection levels in a population. 

Most of the works in mathematical epidemiology deal with the spread of a single virus~\citep{hethcote2000mathematics}. However, it is not unusual to come across settings where multiple virus strains are circulating simultaneously in a population. Such scenarios are far more complicated than single virus spread since those exhibit far richer dynamics \citep{castillo1989epidemiological,santos2015bi,axel2020TAC}. In this paper, we focus on the case where multiple viruses are simultaneously circulating in a population, and these are competitive, i.e., a host can only be infected with one virus at a time. Furthermore, we account for the movement of individuals  across cities even during a pandemic, thus imposing a time-varying graph structure on the interconnection between various individuals.
We adopt the time-varying networked multi-competitive susceptible-infected-susceptible (SIS) model to model the aforementioned aspects. 

A limiting assumption commonly made in disease spread modeling is that contagion occurs due to, and only due to, person-to-person interaction. However, diseases can also spread through other mediums, such as a water distribution network~\citep{Vermeulen2015,la2020first}, and infected surfaces on a public transit network~\citep{hertzberg2018behaviors}. To overcome this shortcoming, a networked susceptible-infected-water-susceptible  (SIWS)  model was recently proposed~\citep{pare2022multi,axel2020TAC,cui2022discrete}. However, existing SIWS models do not account for \emph{time-varying networks} (interconnection between individuals), nor do they provide a sufficient condition for exponential eradication of a virus even when the graph is time-invariant. In light of this observation, we propose a  discrete-time time-varying multi-competitive layered networked SIWS model that also accounts for time-varying graphs.  Our contributions are as follows:
\begin{itemize}[leftmargin=*]
    \item A sufficient condition for global exponential eradication of a virus when graphs are fixed (Theorem~\ref{thm:global:DFE:TI}).
    \item For time-varying graphs, we provide a sufficient condition for global exponential eradication of a virus when:
        \begin{enumerate}
            \item  interaction among individuals is symmetric, and the virus 
            is homogeneous  (same healing and infection rates) (Theorem~\ref{thm:homogenous:spread}); and
            \item  interaction among individuals is asymmetric, and the 
            virus
            is heterogeneous (Theorem~\ref{thm:Main:Result2}).
        \end{enumerate}
\end{itemize}

\textbf{Notations:} Let $\mathbb{R}$ (resp. $\mathbb{Z}_{\geq 0}$) denote the set of real numbers (resp. non-negative integers). We denote the set of positive integers by $\mathbb{Z}_+$. 
Given a matrix $A \in \mathbb{R}^{n \times n}$, $a_{ij}$ denotes the $i^{th}$ row and $j^{th}$ column entry; $\rho(A)$ denotes its spectral radius, and $\lambda_{\min}(A)$ (resp. $\lambda_{\max}(A)$) denotes the minimum (resp. maximum) eigenvalue of $A$ (real).  A diagonal matrix is denoted as $\diag(\cdot)$. The transpose of vector $x \in \mathbb{R}^{n}$ is denoted as $x^\top$ and its average as $\bar{x}:=\frac{1}{n}\sum_{i=1}^n x_i$. Euclidean norms are denoted by $\left\|\cdot\right\|$.  Given a matrix $A$, $A \prec 0$ (resp. $ A\preccurlyeq 0 $) indicates that $A$ is negative definite (resp. negative semidefinite), whereas  $A \succ 0$ (resp. $ A\succcurlyeq 0 $) indicates that $A$ is positive definite (resp. positive semidefinite).
\vspace{-1ex}

\section{Problem Formulation}
\label{sec:2}

We leverage the model proposed in~\citep{cui2022discrete} and generalize it to establish conditions for exponential eradication of a virus. Consider~$m$ competing viruses spreading over a network of~$n$ individuals. Suppose the viruses simultaneously spread over an infrastructure network of~$q$ resource nodes. To avoid the trivial case, we assume $m \geq 2$. The spread of the $r^{th}$ virus, where $r \in [m]$, in individual $i$ can be represented as follows.
\begin{align}
 \scriptsize   \dot{x}_i^{r}(t) =&-\delta_i^{r}x_i^{r}(t) + \Big((1-\textstyle\sum_{\ell=1}^{m}x_i^{\ell}(t)) \times \nonumber  \\ 
 &\big(\textstyle\sum_{j=1}^{n}\beta_{ij}^{r}x_j^r(t) + \textstyle\sum_{j=1}^{q}\beta_{ij}^{wr}w_{j}^{r}(t)\big)\Big),\label{xi}
\end{align}
where $\beta_{ij}^r=\beta_i^ra_{ij}^r$. \seb{The term $\beta_i^r$ (resp. $\delta_i^r$) denotes the infection (resp. healing rate) of individual $i$ for virus~$r$, while $a_{ij}^r\geq 0$ denotes the strength of interconnection between nodes $i$ and $j$ for the spread of virus~$r$. The term $\beta_{ij}^{wr}$ is 
the 
resource-to-individual infection rate  for individual $i$ from resource $j$ for virus~$r$.
Note that $x_i^r(k)$ is an approximation of
the probability of infection with respect to virus~$r$ of individual $i$ at time instant $k$}. 

Viruses can mutate over time, and people move across cities even during the course of a pandemic. Therefore, we 
allow for the healing (resp.) infection rate and the set of neighbors that a node has to vary over time. Thus, \eqref{xi} can be generalized as: 
\begin{align}
 \scriptsize   \dot{x}_i^{r}(t) =&-\delta_i^{r}(t)x_i^{r}(t) + \Big((1-\textstyle\sum_{\ell=1}^{m}x_i^{\ell}(t)) \times \nonumber  \\ 
 &\big(\textstyle\sum_{j=1}^{n}\beta_{ij}(t)^{r}x_j^r(t) + \textstyle\sum_{j=1}^{q}\beta_{ij}^{wr}(t)w_{j}^{r}(t)\big)\Big),\label{xi:tv}
\end{align}
\normalsize
where $\beta_{ij}(t)^r=\beta_i(t)^r a_{ij}(t)^r$, and the concentration of the $r^{th}$ virus in the $j^{th}$ resource node is described as: 
\begin{align}
 \dot w_j^r &{=} {-}\delta^{wr}_{j} w_j^r  {+} \textstyle\sum_{\ell = 1 }^q
        \alpha_{\ell j}^r w_\ell^k {-} w_j^r \textstyle\sum_{\ell = 1 }^q  
    \alpha_{j\ell}^r {+} \textstyle\sum_{\ell=1}^n c^{wr}_{j\ell}(t) x_\ell^r, 
    \label{wj}
 \end{align}
 \normalsize
\seb{
where
$\delta_{j}^{wr}$ denotes the healing rate of resource node~$j$ with respect to virus~$r$; $\alpha_{j\ell}^r$ denotes the  resource-to-resource infection rate  for resource node $\ell$ from resource node $j$; 
and
$c^{wr}_{j\ell}$ denotes the  individual-to-resource infection rate  for resource node $
j$ from individual $\ell$.}

\seb{The spread of the $m$ viruses over a possibly time-varying population network and an infrastructure network can be represented using a time-varying graph. Specifically, we define a multi-layer graph $\mathcal G(k)$ with $m$ layers, where the vertices correspond to  individuals and the shared resource nodes, and layer $r$ 
is the contact graph for the spread of virus~$r$ at time instant $k$, with $r \in [m]$. More precisely, there exists  a directed edge from node $j$ to node $i$ in layer $r$, if individual $j$ (resp. shared resource $\ell$, with $\ell \in [q]$) can infect individual $i$ (resp. shared resource $\ell$) with virus~$r$. For ease of exposition, we define the following sets: $E^r(k)= \{(i,j)\mid i, j \in [n], a_{ji}^r(k) >0\}$; $E_w^r =\{(\ell,j)\mid \ell, j \in [q], a_{\ell j}^r >0 \}$; $E_c^r =\{(j, \ell) \mid \ell \in [n],j \in [q], c_{j\ell}^{wr}(k)>0\}$; and $E_b^r=\{(i,j)\mid i \in [n], j \in [q], \beta_{ij}^{wr}(k)>0\}$. Finally, we define $\mathcal E^r(k)=E^r(k)\cup E_w^r\cup E_c^r(k)\cup E_b^r(k)$. Therefore, layer $r$ of graph $\mathcal G$ at time $k$, denoted by $\mathcal G^r(k)$ is as follows: $\mathcal G^r(k)=(V, \mathcal E^r(k))$, where $\lvert V \rvert = n+q$.} 

Disease outbreaks are often recorded in epidemiological reports that are compiled 
per day \citep{whoCoronavirus,snow1855mode} or per week.  
Thus, the continuous-time spread process is sampled at discrete time intervals. Said sampling of the system behavior
leads to the need for a discrete-time SIWS model. 
The model is
obtained by applying Euler's method \citep{atkinson2008introduction} to \eqref{xi:tv} and~\eqref{wj}, 
\begin{align} 
   & x_{i}^r(k{+}1) = x_{i}^r(k) {+} h\big({-}\delta_i^{r}(k)x_i^{r}(k) {+} (1{-}\textstyle\sum_{\ell=1}^{m}x_i^{\ell}(k)) \times \nonumber  \\ 
 & \qquad \big(\textstyle\sum_{j=1}^{n}\beta_{ij}^{r}(k)x_j^r(t) + \textstyle\sum_{j=1}^{q}\beta_{ij}^{wr}(k)w_{j}^{r}(t)\big)\big) \label{xi:dt}\\
&  w_{j}^r(k+1) = w_{j}^r(k) + h\big(-\delta^{wr}_{j} w_j^r  + \textstyle\sum_{\ell = 1 }^q
        \alpha_{\ell j}^r w_\ell^r  \nonumber \\
        & \qquad -w_j^r \textstyle\sum_{\ell = 1 }^q  
    \alpha_{j\ell}^r + \textstyle\sum_{\ell=1}^n c^{wr}_{j\ell}(k) x_\ell^r\big),
    \label{wj:dt}
\end{align}
\normalsize
\noindent where $h$ is the sampling parameter ($h >0$). In vector form, equations~\eqref{xi:dt} and~\eqref{wj:dt} can be written as follows: 
\begin{align}
    x^r(k+1)=&x^r(k) + h\big(((I-\textstyle\sum_{\ell=1}^{m}X^\ell)B^r-D^r)x^r(k)) \label{xr} \nonumber \\
    &~~~~(I-\textstyle\sum_{\ell=1}^{m}X^\ell)B_w^rw^r(k)\big) \\
     w^r(k{+}1){=}&w^r(k) {+} h({-}D_w^r w^r(k)  {+} A_w^r w^r(k) {+} C_w^r(k) x^r(k)). \label{wr}
\end{align}
\normalsize
\noindent System~\eqref{xr}-\eqref{wr} 
can be more compactly written using 
\begin{align}
\nonumber
 z^r(k) 
 &\coloneqq
 \begin{bmatrix}
 x^r(k)\\
 w^r(k)
 \end{bmatrix}
 ,
 \,\,
 X(z^r(k))  
 \coloneqq
 \begin{bmatrix}
 \diag(x^r(k)) & 0 \\
 0 & 0 
 \end{bmatrix},
 \\
 \normalsize B_f^k(t) 
 &\coloneqq
 \begin{bmatrix}
 B^r(k) & B_w^r(k) \\
 C_w^r(k) & A_w - \diag (A_w) 
 \end{bmatrix}
 ,
 \text{ and}\label{eq:z}
 \\
 D_f^r(k) 
 &\coloneqq
 \begin{bmatrix}
 D^r(k) & 0 \\
 0 & D_w^r  - \diag (A_w)
 \end{bmatrix}
\nonumber
 .
\end{align}
\normalsize 
\noindent Hence, \eqref{xr}-\eqref{wr} can be rewritten as: 
\begin{equation} \label{eq:full}
   z^r(k+1) {=}z^r(k)+ h\big{(} {-} D_f^r(k) {+} (I {-} \textstyle\sum_{\ell=1}^m X(z^\ell)) B_f^r(k) \big{)} z^r(k),
\end{equation}
\normalsize with $r=1,2,\hdots,m$.
\begin{remark}\label{rem:cui}
By setting $A_w=\mathbf{0}$, and $a_{ij}(k)=a_{ij}$ for all $k \in \mathbb{Z}_{\geq 0}$, \eqref{eq:full} coincides with the model in \citep{cui2022discrete}. 
\end{remark}

\begin{remark}
By setting $w^r(k)=\textbf{0}$ for $r=1,2,\hdots, q$, and $m=1$ \eqref{eq:full} collapses to the standard discrete-time time-varying networked SIS model studied in \citep{gracy2020analysis}.
\end{remark}
This paper deals with the stability analysis of the healthy state for the time-varying model in~\eqref{eq:full} and its time-invariant version. To this end, we need the following:
\begin{align} \label{jsr}
  M_f^r(k): = &I-hD_f^r(k)+hB_f^r(k)  \\  
      \hat{M}_f^r(k): =& I-hD_f^r(k)+hB_f^r(k)-h\textstyle\sum_{\ell=1}^m X(z^\ell)B_f^r(k) \nonumber
    \end{align}
\normalsize
Observe that the matrix $M_f^r(k)$ is the state matrix obtained by linearizing the dynamics of virus~$r$ around the eradicated state of virus~$r$ ($x^r(k) = 0$).

\section{Exponential eradication of a virus: Time-Invariant Case}

Let us first consider the case where the interconnection graph is time-invariant, \seb{i.e., $\mathcal G^r(k)= \mathcal G^r$ for all $k \in \mathbb{Z}_{\geq 0}$. Then the spread dynamics is 
as follows:}  
\begin{equation} \label{eq:full:TI}
 \hspace{-0.2cm}    z^r(k{+}1) {=}z^r(k){+} h\big( {-} D_f^r {+} (I {-} \textstyle\sum_{\ell=1}^m X(z^\ell)) B_f^r \big) z^r(k). \hspace{-0.1cm}
\end{equation}

We assume the following for \eqref{eq:full:TI} to be well-defined.
\begin{assm}\label{assm:init:state}
For all $i \in [n]$, $\textstyle\sum_{\ell=1}^{m}x_i^{\ell}(0) \in [0,1]$.
\end{assm}
\begin{assm}\label{assm:posrates}
For all 
$i,j \in [n]$, $r \in [m]$ $\delta_i^r>0$, $\beta_{ij}^r\geq 0$,  $\beta_{ij}^{wr}\geq 0$. For all $r \in [m]$, $i \in [n]$, and $j \in [m]$, $\delta_j^{rw}>0$ and $c_{ij}^{rw}\geq 0$ with at least one $i$ such that $c_{ij}^{rw}> 0$.
\end{assm}

\begin{assm}\label{assm:bound:water:infection}
For all 
$r \in [m]$, $i \in [n]$ and $j \in [q]$, \mbox{$w_j^r(0){\geq} 0$} and \mbox{$w_j^r(0){\leq} w_{max}^r$}, and \mbox{${\textstyle\sum_{\ell=1}^{n}c_{j\ell}^{wr}}/{\delta_{j}^{wr}} \in [0, w_{max}^r]$}.
\end{assm}

\begin{assm}\label{assm:boundsonsum}
 For all $i \in [n]$ (resp. $j \in [q]$), 
 $r \in [m]$,  $h\delta_i^r\in [0,1]$ (resp. $h\delta_j^r\in [0,1]$). Furthermore, $h\textstyle\sum_{\ell=1}^{m} \big(\textstyle\sum_{p=1}^n\beta_{ip}^\ell+ \textstyle\sum_{p=1}^n\beta_{ip}^{w\ell}w^\ell_{max}\big) \in [0,1]$.
\end{assm}

Define $\mathcal D^r =\{z^r(k) =[x^r(k)^\top, w^r(k)^\top]^\top \mid x^r(k) \in [0,1]^n, w^r(k) \in [0, w^r_{max}]^q\}$. Virus~$r$ is eradicated if $z^r(k)=0$. The discrete-time multi-competitive layered networked SIWS model is in the disease-free equilibrium (DFE) if $z^r(k)=0$,  $\forall r \in [m]$.

The following lemma guarantees that the set $\mathcal D^r$ is positively invariant for system~\eqref{eq:full:TI}.
\begin{lem}{\citep[Lemma~1]{cui2022discrete}}
Consider~\eqref{eq:full:TI}, and let Assumptions~\ref{assm:init:state}-\ref{assm:boundsonsum} hold. Then, $x_i^r(k) \in [0,1]$ for all $i \in [n]$, and $z^r_j(k) \in [0, w^r_{max}]$ for all $j \in [q]$, for all $k \in \mathbb{Z}_{\geq 0}$.
\end{lem}
Recall that $x_i^r(k)$ is an approximation of
the probability of infection for virus~$r$ of individual $i$, whereas $z_j^r(k)$ is the concentration of virus~$r$ in resource~$j$; hence, if the states were to take values outside those in set $\mathcal D^r$, then those states would not correspond to physical reality.
Hence, for our subsequent stability results, we prove the system's eradicated state of virus $r$ is stable with the domain of attraction~$\mathcal D^r$, which is equivalent to global stability for this system. \seb{In particular, if the system's eradicated states are stable with the domain of attraction~$\mathcal D^r$ for all $r \in [m]$, then the DFE is globally exponentially stable.} Next, we provide a sufficient condition for the eradication of virus~$r$. 

\begin{theorem}\label{thm:global:DFE:TI}
Let Assumptions~\ref{assm:init:state}-\ref{assm:boundsonsum} hold, and consider system~\eqref{eq:full:TI}. If $\rho(M_f^r)<1$,  with $r \in [m]$, then the eradicated state of virus $r$ is exponentially stable, with domain of attraction~$\mathcal D^r$. 
\end{theorem}

\textit{Proof:} By Assumption~\ref{assm:boundsonsum}, we have that, for each $i \in [n]$ (resp. $j \in [q]$) $h\delta_i^r\in [0,1]$ (resp. $h\delta_j^r\in [0,1]$), which implies that the matrix $I-D_f^r$ is nonnegative. Therefore, 
noting that $h>0$, and since Assumption~\ref{assm:posrates} implies that the matrix $B_f^r$ is nonnegative, we have that $M_f^r$ is nonnegative.\\
By assumption, $\rho(M_f^r)<1$. Hence, from \citep[Prop.~1]{rantzer2011distributed} it follows that there exists a positive diagonal matrix $P^r$ such that $(M_f^r)^\top P^r M_f^r-P^r \preccurlyeq 0$. Consider the Lyapunov function candidate $V(z^r)=(z^r)^\top P^rz^r$. Since $P^r > 0$, it follows that $V(z^r)>0$ for all $z^r \neq 0$. Since $P^r > 0$, it is also symmetric. Therefore, by applying the Rayleigh-Ritz Theorem (RRT) \citep{horn2012matrix}. Thus, $\lambda_{\min}(P^r)I \leq P^r \leq \lambda_{\max}(P^r)I$, and 
\begin{align}\label{eq:positiveconstants}
\lambda_{\min}(P^r)\left\|z\right\|^{2} \leq V(z^r) \leq \lambda_{\max}(P^r)\left\|z\right\|^{2}.
\end{align}

Observe that since $P^r >0$, all its eigenvalues are positive; hence, $\lambda_{\min}(P^r)>0$ and $\lambda_{\max}(P^r)>0$. Therefore,~\eqref{eq:positiveconstants} implies that the constants bounding the Lyapunov function candidate are strictly positive. \\
Define $\Delta V(z^r): = V(z^r(k+1)) - V(z^r(k))$. Hence, for all $z^r\neq 0$, we have the following: 

\vspace{-3ex}
\scriptsize
\begin{align}
    & \Delta V(z^r)    = z^r(k+1)^\top P^r z^r(k+1) - z^r(k)^\top P^r z^r(k) \nonumber \\
    =& z^r(k)^\top \hat{M}_f^r{}^\top P^r \hat{M}_f^r z^r(k) - z^r(k)^\top P^r z^r(k) \nonumber \\
    =& z^r(k)^\top \big((M_f^r -h\textstyle\sum_{\ell=1}^m X(z^\ell)B_f^r)^\top P^r (M_f^r  \nonumber \\ &~~~~~~~~~~-h\textstyle\sum_{\ell=1}^m X(z^\ell)B_f^r)\big) z^r(k) 
    -z^r(k)^\top P^r z^r(k) \nonumber \\
    = & (z^r)^\top \hat{M}_f^r{}^\top P^r M_f^r z^r - (z^r)^\top P^r z^r \nonumber \\ &-2h(z^r)^\top\textstyle\sum_{\ell=1}^m X(z^\ell)B_f^r P^r M_f^r z^r \nonumber \\
    & ~+h^2(z^r)^\top \Big((\textstyle\sum_{\ell=1}^m X(z^\ell)B_f^r)^\top P^r \textstyle\sum_{\ell=1}^m X(z^\ell)B_f^r\Big) z^r .
    \label{eq:lyap:diff}
\end{align}
\normalsize 
Observe that \scriptsize
\begin{align}
   &-2h(z^r)^\top\textstyle\sum_{\ell=1}^m X(z^\ell)B_f^rP^rM_f^rz^r +h^2(z^r)^\top \times \nonumber \\
   &\qquad \qquad \Big((\textstyle\sum_{\ell=1}^m X(z^\ell)B_f^r)^\top P^r \textstyle\sum_{\ell=1}^m X(z^\ell)B_f^r\Big) z^r \nonumber \\
   &\leq (z^r)^\top \Big(h^2 (B_f^r)^\top \textstyle\sum_{\ell=1}^m X(z^\ell)P^r \textstyle\sum_{\ell=1}^m X(z^\ell)B_f^r \nonumber \\
   &~~~~~~~~~~~~-2h^2(B_f^r)^\top P^r \textstyle\sum_{\ell=1}^m X(z^\ell)B_f^r\Big)z^r \label{ineq:1}\\
   &\leq (z^r)^\top h^2 \Big((B_f^r)^\top \textstyle\sum_{\ell=1}^m X(z^\ell)P^r \textstyle\sum_{\ell=1}^m X(z^\ell)B_f^r \nonumber \\
   &~~~~~~~~~~~~~~~~~~-(B_f^r)^\top P^r \textstyle\sum_{\ell=1}^m X(z^\ell)B_f^r\Big)z^r \label{ineq:2}\\
   & =-{z^r}^\top h^2 \Big((B_f^r)^\top (I-\textstyle\sum_{\ell=1}^m X(z^\ell)P^r \textstyle\sum_{\ell=1}^m X(z^\ell) B_f^r\Big)z^r \leq 0, \label{ineq:3}
\end{align} \normalsize 
where inequality~\eqref{ineq:1} comes from noting that i)  due to  Assumption~\ref{assm:posrates} the matrix $B_f^r$ is nonnegative, and ii)  due to  Assumption~\ref{assm:boundsonsum}, the matrix $(I-hD_f^r)$ is nonnegative. 
Consequently, the term $-2h(z^r)^\top (I-D_f^r)P^r\textstyle\sum_{\ell=1}^m X(z^\ell)B_f^r(z^r)$ is nonpositive. Inequality~\eqref{ineq:2} is a consequence of Assumption~\ref{assm:boundsonsum}, whereas inequality~\eqref{ineq:3} follows by extending the argument in \citep[Lemma~6]{axel2020TAC} 
to
the $m$-virus case. Therefore, from~\eqref{eq:lyap:diff}, it follows that 
\begin{align}
   \Delta V(z^r) \leq  (z^r)^\top \big( {M_f^r}^\top P^r M_f^r  -  P^r \big)z^r. \label{ineq:lyap:bound}
\end{align} 
Since, as seen above, $(M_f^r)^\top P^r M_f^r-P^r$ is negative definite, it follows that $(M_f^r)^\top P^r M_f^r-P^r$ is symmetric; hence, its spectrum is real, and all its eigenvalues are negative. Therefore, by RRT, we have
\begin{align}
     \Delta V(z^r)  \leq -\lambda_{\min}(P^r -(M_f^r)^\top P^r M_f^r)\left\|z\right\|^{2}, \label{constant3}
\end{align}
where $\lambda_{\min}(P^r {-}(M_f^r)^\top P^r M_f^r)>0$. From~\eqref{eq:positiveconstants} and~\eqref{constant3}, we have that  there exists positive constants, $\lambda_{\min}(P^r)$, $\lambda_{\max}(P^r)$, and $\lambda_{\min}(P^r {-}(M_f^r)^\top P^r M_f^r)$, such that for~\mbox{$z {\neq} 0$},  
\begin{align}
\lambda_{\min}(P^r)\left\|z\right\|^{2} \leq V(z^r) \leq \lambda_{\max}(P^r)\left\|z\right\|^{2}, \label{a1}\\
     \Delta V(z^r)  \leq -\lambda_{\min}(P^r -(M_f^r)^\top P^r M_f^r)\left\|z\right\|^{2}.\label{a2}
\end{align}

The result then follows as a direct consequence of \citep[Section~5.9 Theorem.~28]{vidyasagar2002nonlinear}.~\qed

The following result is immediate.
\begin{cor}\label{thm:DFE:exp:stable}
Consider system~\eqref{eq:full} under Assumptions~\ref{assm:init:state}-\ref{assm:boundsonsum}. If $\rho(M_f^r)<1$, for all $r \in [m]$, then the DFE is globally exponentially stable.
\end{cor}


Corollary~\ref{thm:DFE:exp:stable} provides guarantees for exponential convergence to the DFE, while \citep[Theorem~10]{cui2022discrete} only provides asymptotic guarantees for the same. Moreover, Corollary~\ref{thm:DFE:exp:stable}, unlike \citep[Theorem~10]{cui2022discrete}, does not require the graph to be strongly connected. On the other hand, \citep[Theorem~10]{cui2022discrete} relaxes the condition  
on the spectral radius of $M_f^r$ in Corollary~\ref{thm:DFE:exp:stable} and yet achieves  convergence, albeit asymptotic, to the healthy state; thus guaranteeing eradication of viruses for a larger range of model parameters. The term $\rho(M_f^r)$ can be interpreted as the reproduction number for virus $r$. Define $M^r: = I-hD+hB$; the term $\rho(M^r)$ denotes the reproduction number for virus~$r$ \textit{assuming there is no infrastructure network}. It is natural to explore the relation between $\rho(M_f^r)$ and $\rho(M^r)$. To this end, we need the following assumption and proposition.
\begin{assm}\label{assm:irreducibility}
The matrix $B_f^r$ is irreducible for $r \in [m]$.
\end{assm}
\begin{proposition}\label{prop:reprod:number}
Consider system~\eqref{eq:full} under Assumptions~\ref{assm:posrates}, \ref{assm:boundsonsum}, and~\ref{assm:irreducibility}. The reproduction number of the  multi-virus SIS network with an infrastructure network is greater than the reproduction number of the  multi-virus SIS network without the infrastructure network, i.e., $\rho(M_f^r)>\rho(M^r)$.
\end{proposition}
\textit{Proof:} Consider the matrix $M_f^r$ and notice that, due to Assumption~\ref{assm:irreducibility}, it is irreducible, whereas due to Assumptions~\ref{assm:posrates} and~\ref{assm:boundsonsum} it is nonnegative. Furthermore, it can be expressed as follows: 
\begin{equation}
    M_f^r = \begin{bmatrix}
    M^r && hB_w^r \\
    hC_w^r && I-hD_w^r+hC_w^r \nonumber
    \end{bmatrix}.
\end{equation}
\normalsize 
Note that $M^r$ is a principal square submatrix of $M_f^r$. Therefore, from \citep[Lemma 2.6]{varga}, it follows that $\rho(M_f^r) > \rho(M^r)$.\qed

Proposition~\ref{prop:reprod:number} implies that eradicating a virus in the population network does not necessarily imply eradication of 
said virus in the layered network; this further underscores the challenges of combating epidemics that spread through multiple mediums. 

\section{Exponential eradication of a virus: Time-Varying Case}

This section studies the case where the population network is time-varying, \seb{i.e, we allow for $\mathcal G^r(k_0) \neq \mathcal G^r(k_1)$ for any $k_0 \neq k_1 \in \mathbb{Z}_{\geq 0}$.} We rely on the model in~\eqref{eq:full}. Before proceeding with the analysis, we need the following assumptions to ensure that~\eqref{eq:full} is well-defined. 
\begin{assm}\label{assm:posrates:TV}
For all $k \in \mathbb{Z}_{\geq 0}$, $i,j \in [n]$, $r \in [m]$ $\delta_i^r(k)>0$, $\beta_{ij}^r(k)\geq 0$,  $\beta_{ij}^{wr}(k)\geq 0$. For all $r \in [m]$, $i \in [n]$, and $j \in [m]$, $\delta_j^{rw}>0$ and $c_{ij}^{rw}\geq 0$ with at least one $i$ such that $c_{ij}^{rw}> 0$.
\end{assm}

\begin{assm}\label{assm:bound:water:infection:TV}
For all $k \in \mathbb{Z}_{\geq 0}$, $r \in [m]$, $i \in [n]$ and $j \in [q]$, $w_j^r(0)\geq 0$ and $w_j^r(0)\leq w_{max}^r$. Furthermore, ${\textstyle\sum_{\ell=1}^{n}c_{j\ell}^{wr}(k)}/{\delta_{j}^{wr}(k)} \in [0, w_{max}^r]$.
\end{assm}

\begin{assm}\label{assm:boundsonsum:TV}
 For all $i \in [n]$ (resp. $j \in [q]$), $k \in \mathbb{Z}_{\geq 0}$ and $r \in [m]$,  $h\delta_i^r(k)\in [0,1]$ (resp. $h\delta_j^r(k)\in [0,1]$). Furthermore, $h\textstyle\sum_{\ell=1}^{m} \big(\textstyle\sum_{p=1}^n\beta_{ip}^\ell(k)+ \textstyle\sum_{p=1}^n\beta_{ip}^{w\ell}(k)w^\ell_{max}\big) \in [0,1]$.
\end{assm}

Assumptions~\ref{assm:posrates:TV},~\ref{assm:bound:water:infection:TV}, and \ref{assm:boundsonsum:TV} imply Assumptions~\ref{assm:posrates},~\ref{assm:bound:water:infection}, and \ref{assm:boundsonsum}, respectively. The converse, however, is false. The following lemma establishes positive invariance of the set $\mathcal D^r$ for~\eqref{eq:full}.

\begin{lem}\label{lem:pos:invaraince:tv}{\citep[Lemma~4]{cui2022discrete}}
Let Assumptions~\ref{assm:init:state},~\ref{assm:posrates:TV}-\ref{assm:boundsonsum:TV} hold and consider~\eqref{eq:full}. Then $x_i^r(k) \in [0,1]$, $\forall i \in [n]$, and $z^r_j(k) \in [0, w^r_{max}]$,  $\forall j \in [q]$, $\forall k \in \mathbb{Z}_{\geq 0}$.
\end{lem}

\subsection{Homogeneous spread, symmetric undirected graphs}

We focus on homogeneous virus spread (i.e., the infection rate for a virus is the same for every individual) in the layered network. The following theorem identifies a sufficient condition for the exponential eradication of a virus, irrespective of the initial infection levels in the network of individuals and in the network of shared resources, for the 
virus.
\begin{theorem}
\label{thm:homogenous:spread}
Consider system~\eqref{eq:full} under Assumptions~\ref{assm:init:state},~\ref{assm:posrates:TV}-\ref{assm:boundsonsum:TV}. Suppose that for all $k \in \mathbb{Z}_{\geq 0}$
\begin{enumerate}[label=\roman*)]
    \item $\beta_i^r(k) {=} \beta^r(k)$  $\forall i \in [n]$ \cub{(Homogeneous infection rate)};
     \item $\delta_i^r(k) {=} \delta^r(k)$ $\forall i \in [n]$ \cub{(Homogeneous healing rate)};
     \item $A^r(k){=}A^r(k)^\top$ \cub{(Symmetric social interactions)}; and
     \item $B_w^r(k){=}C_w^r(k)^\top$ \cu{(Sym. infrastructure interactions)}.
\end{enumerate}
If $\sup_{k \in \mathbb{Z}_{\geq 0}} \rho(M_f^r(k))<1$, \seb{where $r \in [m]$,} then the eradicated state of virus~$r$ is exponentially stable with a domain of attraction~$\mathcal D^r$. 
\end{theorem}
\textit{Proof:} Consider the Lyapunov function candidate $V(z^r,k) = \frac{1}{2}z^r(k)^\top z^r(k)$. It is immediate that $V(z^r,k) >0$ for all $k$ and $z^r(k) \neq 0$. Define $\Delta V(z^r,k): = V(z^r(k+1)) - V(z^r(k))$. Hence, for all $z^r\neq 0$, we have the following: \scriptsize
\begin{align}
    &\Delta V(z^r)    = \frac{1}{2}\big(z^r(k+1)^\top z^r(k+1) - z^r(k)^\top z^r(k)\big) \nonumber \\
    =& \frac{1}{2}\big(z^\top(M_f^r(k)-\textstyle\sum_{\ell=1}^{m}Z^\ell B_f^\ell)^\top (M_f^r(k)-\textstyle\sum_{\ell=1}^{m}Z^\ell B_f^\ell)z^r-(z^r)^\top z^r\big) \nonumber \\
     =& \frac{1}{2}\big((z^r)^\top (M_f^r(k)^\top M_f^r -hM_f^r(k)^\top \textstyle\sum_{\ell=1}^{m}Z^\ell B_f^\ell \nonumber \\
     &~-hB_f^r(k)^\top \textstyle\sum_{\ell=1}^{m}Z^\ell M_f^r(k) 
     +h^2B_f^r(k)^\top \textstyle\sum_{\ell=1}^{m}Z^\ell \textstyle\sum_{\ell=1}^{m}Z^\ell B_f^r(k))z^r \nonumber \\&~~~-(z^r)^\top z^r\big). \label{eq:key:lyap:homogenous:1}
\end{align}
\normalsize 
Observe that \scriptsize
\begin{align}
 &(z^r)^\top\big(h^2 B_f^r(k)^\top \textstyle\sum_{\ell=1}^{m}Z^\ell \textstyle\sum_{\ell=1}^{m}Z^\ell B_f^r(k) -2h M_f^r(k)^\top \textstyle\sum_{\ell=1}^{m}Z^\ell B_f^r(k) \big) z  \nonumber \\
 &=(z^r)^\top \big( h^2 B_f^r(k)^\top \textstyle\sum_{\ell=1}^{m}Z^\ell \textstyle\sum_{\ell=1}^{m}Z^\ell B_f^r(k) \nonumber \\
 &~~~-2h^2B_f^r(k)^\top\textstyle\sum_{\ell=1}^{m}Z^\ell B_f^r(k) \nonumber \\
 &~~~-2h (I-hD_f^r(k))\textstyle\sum_{\ell=1}^{m}Z^\ell B_f^r(k)\big)z^r \nonumber \\
 &\leq (z^r)^\top\big(h^2 B_f^r(k)\textstyle\sum_{\ell=1}^{m}Z^\ell \textstyle\sum_{\ell=1}^{m}Z^\ell B_f^r(k) -2h^2B_f^r(k)^\top\textstyle\sum_{\ell=1}^{m}Z^\ell B_f^r(k) \big)z^r  \label{ineq:lyap:TV:1} \\
 & \leq (z^r)^\top\big(h^2 B_f^r(k)\textstyle\sum_{\ell=1}^{m}Z^\ell \textstyle\sum_{\ell=1}^{m}Z^\ell B_f^r(k) -h^2B_f^r(k)^\top\textstyle\sum_{\ell=1}^{m}Z^\ell B_f^r(k) \big)z^r  \label{ineq:lyap:TV:2} \\
 & =-(z^r)^\top\big(h^2B_f^r(k)^\top \textstyle\sum_{\ell=1}^{m}Z^\ell (I-\textstyle\sum_{\ell=1}^{m}Z^\ell)B_f^r(k)\big)z^r \nonumber \\
 &\leq 0,\label{ineq:lyap:TV:3}
\end{align}
\normalsize
where~\eqref{ineq:lyap:TV:1} follows by noting that i) due to Assumption~\ref{assm:boundsonsum:TV} the matrix $(I-hD_f^r(k))$ is nonnegative; and ii) due to Assumption~\ref{assm:posrates:TV}, the matrix $B_f^r(k)$ is nonnegative; thus, implying that $-(z^r)^\top 2h (I-hD_f^r(k))\textstyle\sum_{\ell=1}^{m}Z^\ell B_f^r(k)z^r \leq 0$. Inequality~\eqref{ineq:lyap:TV:2} is a direct consequence of Lemma~\ref{lem:pos:invaraince:tv} and Assumption~\ref{assm:posrates:TV}, whereas inequality~\eqref{ineq:lyap:TV:3} can be obtained by extending the claim in \citep[Lemma~6]{axel2020TAC} for $m$ arbitrary, but finite, viruses. Plugging~\eqref{ineq:lyap:TV:3} into~\eqref{eq:key:lyap:homogenous:1} yields the following: 
\begin{align}
 \Delta V(z^r, k)   &\leq \frac{1}{2}(z^r)^\top\big(M_f^r(k)^\top M_f^r(k) -I\big)  z^r \label{ineq:lyap:diff:homogenous:TV} 
\end{align}
\normalsize
It follows from the theorem 
assumptions
that $M_f^r(k)$ is symmetric for all $k$, which implies that
\begin{enumerate}[label=\roman*),leftmargin=*]
    \item $\lambda_{\max}(M_f^r(k)^\top M_f^r(k)) = \lambda_{\max}(M_f^r(k))^2$; and
    \item the spectrum of $M_f^r(k)$ is real.
\end{enumerate}
Statement~ii) implies that, for all $k$, $\rho(M_f^r(k)) {=} \lambda_{\max}(M_f^r(k))$.  Therefore, since, by assumption,  $\sup_{k \in  \mathbb{Z}_{\geq 0}}\rho(M_f^r(k))<1$, statement~i) and the definition of supremum together imply that $\lambda_{\max}(M_f^r(k)^\top M_f^r(k)) <1$. Applying Weyl's inequalities \citep[Corollary 4.3.15]{horn2012matrix} to $M(k)^\top M(k)-I$, we obtain, for $i =1,2, \hdots n+q$, 
$\lambda_{i}(M_f^r(k)^\top M_f^r(k)-I) \leq \lambda_{i}(M_f^r(k)^\top M_f^r(k)) -1$. Since, for every $k \in \mathbb{Z}_{\geq 0}$, 
$\lambda_{\max}(M_f^r(k)^\top M_f^r(k)) <1$, it follows that, for each $k \in \mathbb{Z}_{\geq 0}$, $\lambda_i M_f^r(k)^\top M_f^r(k)-I) < 0$ for $i \in [n+q]$. Plugging  $\lambda_i M_f^r(k)^\top M_f^r(k)-I) < 0$ back into~\eqref{ineq:lyap:diff:homogenous:TV} yields: \break $(z^r)^\top\big(M_f^r(k)^\top M_f^r(k) -I\big)  z^r <0$ for $z^r \neq 0$ and $k \in \mathbb{Z}_{\geq 0}$. Hence, it follows that, for $z^r \neq 0$ and $k \in \mathbb{Z}_{\geq 0}$, $\Delta V(z^r,k)<0$. Exponential eradication of virus~$r$ with a domain of attraction $\mathcal D^r$, then, follows from \citep[Theorem~28, Section~5.9]{vidyasagar2002nonlinear}. \qed

\subsection{Directed networks and Heterogeneous spread}

We have the following result.
\begin{theorem}\label{thm:Main:Result2}
Let Assumptions~\ref{assm:init:state},~\ref{assm:posrates:TV}-\ref{assm:boundsonsum:TV} hold and consider system~\eqref{eq:full}. Assume  $\exists \  \alpha_1 >0, L \in \mathbb{R}_+, \kappa \in \mathbb{R}_+$, such that
\begin{enumerate}[label=\roman*),leftmargin=*]
    \item  $\sup_{k \in \mathbb{Z}_{\geq 0}} \rho(M_f^r(k)) \leq \alpha_1<1$;
    \item  $\forall k \in \mathbb{Z}_{\geq 0}$ $\lvert|M_f^r(k)\rvert| \leq L$; and
    \item  $\sup_{k \in \mathbb{Z}_{\geq 0}} \lvert|M_f^r(k+1)-M_f^r(k)\rvert|\leq \kappa$.
\end{enumerate}
If $\kappa$ is sufficiently small, then the eradicated state of virus~r is  exponentially stable, with a domain of attraction $\mathcal D^r$.
\end{theorem}

We provide an explicit expression for $\kappa$ later in the proof. The proof of Theorem~\ref{thm:Main:Result2} closely mirrors that of \citep[Theorem~2]{pare2020data}; it can be traced back to the linear work in   \citep{desoer1970slowly,rugh1996linear}. In the interest of completeness, we provide all the details here.

\textit{Proof:} Consider the discrete-time Lyapunov equation: 
\begin{equation} \label{eq:Lyapunov}
    (M_f^r)^\top(k)Q(k+1)M_f^r(k) - Q(k+1) = -I_{n+q} .
\end{equation}
\normalsize 
Observe that $I_{n+q}$ is symmetric and positive definite. Moreover, by assumption $\sup_{k \in \mathbb{Z}_{\geq 0}} \rho(M_f^r(k)) <1$. Therefore, 
the solution to~\eqref{eq:Lyapunov} (say, $Q(k+1)$) exists, is unique and is positive definite for all $k
\in \mathbb{Z}_{\geq 0}$; see \citep[Theorem~23.7]{rugh1996linear}. Furthermore,  from the proof of \citep[Theorem~24.8]{rugh1996linear}, a closed-form expression for the solution is as follows: 
\begin{equation} \label{eq:soln:lyapunov}
Q(k+1) = I_{n+q} +  \textstyle\sum_{j=1}^{\infty}[(M_f^r)^\top(k)]^{j}(M_f^r)^{j}(k). 
\end{equation}
\normalsize 
Consider the Lyapunov function $V(k,z^r)=(z^r)^\top Q(k)z^r$. Given that, for each $k \in \mathbb{Z}_{\geq 0}$, $Q(k)$ is positive definite, it follows that $V(z^r, k)>0$ for all $k \in \mathbb{Z}_{\geq 0}$ and $z^r \neq 0$. The rest of the proof can be broken down into three steps: First, we find a constant $\gamma_1>0$ such that $\gamma_1\lvert|z^r\rvert|^2 \leq V(k, z^r)$ for all $k \in \mathbb{Z}_{\geq 0}$. Second, we find a constant $\gamma_2>0$ such that $V(k, z^r) \leq \gamma_2\lvert|z^r\rvert|^2$ for all $k \in \mathbb{Z}_{\geq 0}$. Finally, we  prove that $\Delta V(k, z^r) <0$ for all $z^r \neq 0$ and $k \in \mathbb{Z}_{\geq 0}$.

\textit{Step 1:} From~\eqref{eq:soln:lyapunov} it is immediate that $Q(k) \geq I$ for all $k$. Therefore, $(z^r)^\top z^r \leq (z^r)^\top Q(k)z^r$, and hence we have for all $k\in \mathbb{Z}_{\geq 0}$: $\lvert|z^r\rvert|^2 \leq V(z^r, k)$

\textit{Step 2:} Our objective here is to find an upper bound on $V(k,x)$, which is independent of $k$. To this end, define $\mu:=\frac{1-\alpha_1}{2}$. Therefore, the assumption $\sup_{k \in \mathbb{Z}_{\geq 0}} \rho(M_f^r(k)) \leq \alpha_1$ implies that $\sup_{k \in \mathbb{Z}_{\geq 0}} \rho(M_f^r(k)) \leq 1-2\mu$. It can be easily verified that $1-\mu >0$. By using Dunford's integral \citep[page 568]{dunford1958linear} with the circle of radius $1-\mu$ as contour, we have the following: 
\begin{align} \label{eq:dunford}
    & M_f^r(k)^P = \frac{1}{2\pi j}\oint\limits_{C}s^P(sI_{n+q}-M_f^r(k))^{-1}ds 
    \nonumber \\
    & \leq \frac{1}{2\pi j}2\pi\lvert s\rvert\max_{\lvert s\rvert =1-\mu}\{s^F(sI_{n+q}-M_f^r(k))^{-1}\}. 
\end{align}
\normalsize
 By taking the norm of both sides of \eqref{eq:dunford}, and evaluating at $\lvert s\rvert = 1-\mu$ one obtains: 
\begin{align} 
 & \lvert|M_f^r(k)^P \rvert| {\leq} (1{-}\mu) \max_{\lvert s\rvert = 1{-}\mu}(\lvert s \rvert^P) \lvert|\max_{\lvert s\rvert = 1{-}\mu}(sI_{n+q}-M_f^r(k))^{{-}1}\rvert| \nonumber \\
  &\lvert|M_f^r(k)^P \rvert|  \leq (1-\mu)^{P+1}  \lvert|\max_{\lvert s\rvert = 1-\mu}(sI_{n+q}-M_f^r(k))^{-1}\rvert|  \nonumber \\
  & \leq (1-\mu)^{P+1}\max_{\lvert s\rvert = 1-\mu}\lvert|(sI_{n+q}-M_f^r(k))^{-1}\rvert| \nonumber \\
    & \leq (1-\mu)^{P+1}\max_{\lvert s\rvert = 1-\mu}\Big\{\frac{\lvert|(sI_{n+q}-M_f^r(k))\rvert|^{n+q-1}}{\lvert \text{det}(sI_{n+q}-M_f^r(k))\rvert}\Big\}, \label{eq:norm:kato}
\end{align}
\normalsize
where \eqref{eq:norm:kato} is due to \cite[Lemma 1]{kato1960estimation}.

From \citep[pg.55]{horn2012matrix} it is clear that, given a $s \in \mathbb{C}$, $\text{det}(sI_{n+q}-M_f^r(k)) = (s-\lambda_{j}(M_f^r(k)))^{n+q}$. Notice that
\begin{align}
\lvert s-\lambda_{j}(M_f^r(k))\rvert &\geq \lvert\lvert s \rvert - \lvert \lambda_{j}(M_f^r(k))\rvert \rvert \label{ineq:2_1}\\
& \geq \lvert \lvert s \rvert - (1-2\mu)\rvert \label{ineq:2_2} \\
& = \mu, \label{ineq:2_3}
\end{align}
\normalsize
where~\eqref{ineq:2_1} follows from the reverse triangle inequality. We obtain inequality~\eqref{ineq:2_2} in view of the following: Recall that $\sup_{k \in \mathbb{Z}_{\geq 0}} \rho(M_f^r(k)) \leq 1-2\mu$. By employing the definition of supremum, it must be that, for every $k$, each pointwise eigenvalue of $M_f^r(k)\leq 1-2\mu$, i.e., $\lvert\lambda_i(M_f^r(k))\rvert\leq 1-2\mu$, where $i=1,2,\hdots n+q$. Equality~\eqref{ineq:2_3} is obtained by evaluating~\eqref{ineq:2_2} at $\lvert s\rvert=1-2\mu$, and, as a result, for $\lvert s \rvert=1-\mu$, $\lvert \text{det}(sI_{n+q}-M_f^r(k))\rvert \geq \mu^{n+q}$.

By assumption there also exists an $L$ such that $\lvert|M_f^r(k)\rvert| \leq L$, for all $k
\in \mathbb{Z}_{\geq 0}$. Consequently, $\lvert|(sI_{n+q}-M_f^r(k))\rvert| \leq (1-\mu + L)$. Therefore, given that $\lvert \text{det}(sI_{n+q}-M_f^r(k))\rvert \geq \mu^{n+q}$, we can rewrite~\eqref{eq:norm:kato} as follows: 
\begin{align}\label{eq:desoer}
\lvert|M_f^r(k)^P \rvert| & \leq \frac{(1-\mu)^{P+1}}{\mu ^{n+q}}(1-\mu + L)^{n+q-1}.
\end{align}
\normalsize 
Define $m_1: = \frac{1-\mu}{\mu^{n+q}}(1-\mu + L)^{n+q-1}$ and $p_1:= (1-\mu)$. Therefore, \eqref{eq:desoer} can be rewritten as:
\begin{align}\label{eq:desoer:1}
\lvert|M_f^r(k)^P \rvert| & \leq m_1p_1^P \hspace{4mm} \forall P, \forall k
\in \mathbb{Z}_{\geq 0}.
\end{align}
\normalsize 
Observe that taking norms on both sides of \eqref{eq:soln:lyapunov}, and taking recourse to   the triangle inequality and the submultiplicativity of matrix norms, we obtain: 
\begin{align} 
\lvert|Q(k+1)\rvert| & \leq 1+ \textstyle\sum_{j=1}^{\infty}m_1^2 p_1^{2j} \leq \frac{m_1^2}{1-p_1^2},\label{desoer:upper:bound2}
\end{align}
\normalsize 
Note that $p_1 <1$, then $p_1^2<1$, which implies~\eqref{desoer:upper:bound2}.  Since $Q(k)$ is symmetric  $\forall k 
   \in \mathbb{Z}_{\geq 0}$, by applying RRQ we have:
   \begin{equation}
   \lambda_{\min}(Q(k))I \leq Q(k) \leq  \lambda_{\max}(Q(k))I, \nonumber
   \end{equation}
   \normalsize 
   which implies
    \begin{align}
  & \lambda_{\min}(Q(k))\lvert|z^r(k) \rvert|^2 \leq z^r(k)^\top Q(k) z^r(k) \nonumber \\
  & \leq  \lambda_{\max}(Q(k))\lvert|z^r(k) \rvert|^2 \leq \lvert|Q(k)\rvert|\cdot \lvert|z^r(k) \rvert|^2 \label{eq:RRQ:1}\\
   & \leq \frac{m^2}{1-p^2} \lvert|z^r(k) \rvert|^2, \label{eq:RRQ:2}
   \end{align}
   \normalsize 
   where~\eqref{eq:RRQ:1} follows from  \cite[Theorem 5.6.9]{horn2012matrix}, and ~\eqref{eq:RRQ:2} is due to ~\eqref{desoer:upper:bound2}. Then, $\forall k \in \mathbb{Z}_{\geq 0}$, 
  \begin{align}
       V(k,z^r) \leq  \frac{m_1^2}{1-p_1^2}\lvert|z^r\rvert|^2.
  \end{align}
  \normalsize 
  \textit{Step 3:} Define $\Delta V(k,z^r): = V(z^r(k+1)) - V(z^r(k))$. Hence, for $z^r \neq 0$, and $\forall k \in \mathbb{Z}_{\geq 0}$, we obtain the following:
\begin{align}
    & \Delta V(k,z^r)    
     =(z^r)^\top(M_f^r(k)^\top Q(k+1) M_f^r(k)-Q(k))z^r \nonumber \\ &~~-2h(z^r)^\top M_f^r(k)^\top Q(k+1)
    \textstyle\sum_{\ell=1}^m X(z^\ell)B_f^rz^r \nonumber \\
    & ~~~+h^2(z^r)^\top B_f^r(k)^\top\textstyle\sum_{\ell=1}^m X(z^\ell) Q(k+1) \textstyle\sum_{\ell=1}^m X(z^\ell)B_f^r 
    z^r. \label{eq:lyap:diff:TV}
\end{align}
\normalsize 
The matrix $M_f^r(k)^\top Q(k+1) M_f^r(k)-Q(k)$ is negative definite. Subtracting two successive instances of~\eqref{eq:Lyapunov} results~in  
\begin{align} \label{eq:diff:1}
M_f^r(k)^\top Q(k{+}1)M_f^r(k) {-} M_f^r(k{-}1)^\top Q(k)M_f^r(k{-}1) \nonumber \\
{=} Q(k{+}1) {-}Q(k)  .
\end{align}
\normalsize
Adding and subtracting $M_f^r(k)^\top Q(k)M_f^r(k)$ to the LHS of~\eqref{eq:diff:1},  and rearranging of terms, leads to 
\begin{align} \label{eq:diff:2}
&M_f^r(k)^\top(Q(k+1)-Q(k))M_f^r(k) - (Q(k+1) -Q(k)) 
 =  \nonumber \\ & M_f^r(k-1)^\top Q(k)M_f^r(k-1) - M_f^r(k)^\top Q(k)M_f^r(k).
\end{align}
\normalsize
In a similar vein, by adding and subtracting $M_f^r(k-1)^\top Q(k)M_f^r(k)$ to the RHS of  \eqref{eq:diff:2}, we obtain
\begin{align} \label{eq:diff:3}
& M_f^r(k)^\top(Q(k+1)-Q(k))M_f^r(k) - (Q(k+1) -Q(k)) \nonumber \\ 
 & =  
- ((M_f^r(k)^\top - M_f^r(k-1)^\top)Q(k)M_f^r(k)  \nonumber \\ &  \ \ \  \ +  M_f^r(k-1)^\top Q(k)(M_f^r(k) - M_f^r(k-1))) .
\end{align}
\normalsize

Define $R_1: = ((M_f^r(k))^\top - (M_f^r(k-1))^\top )Q(k)M_f^r(k) + (M_f^r(k-1))^\top Q(k)(M_f^r(k) - M_f^r(k-1))$. 
As a consequence, we have the following: 
\begin{align} 
&\lvert|R_1\rvert| 
 \leq  \lvert| (M_f^r(k)^\top - M_f^r(k-1)^\top))Q(k)M_f^r(k) \rvert| + \nonumber \\ 
 & \lvert|M^\top(k-1)Q(k)(M(k) - M(k-1))\rvert| \label{tri:ineq}\\
& \leq \lvert|(M_f^r(k)^\top - M_f^r(k-1)^\top)\rvert|\cdot \lvert|Q(k)\rvert|\cdot \lvert|M_f^r(k)\rvert| \nonumber \\
&{+} \lvert|M_f^r(k{-}1)^\top\rvert|\cdot\lvert|Q(k)\rvert|\cdot \lvert|M_f^r(k)^\top {-} M_f^r(k{-}1)^\top)\rvert|. \label{sub:multi}
\end{align} 
\normalsize
\noindent Note that inequality~\eqref{tri:ineq} comes from the triangle inequality of matrix norms, while inequality~\eqref{sub:multi} follows from the submultiplicativity of matrix norms.

Since, for all $k \in \mathbb{Z}_{\geq 0}$, i) by assumption, there exists 
$\kappa$ such that
$\lvert|M_f^r(k+1)-M_f^r(k)\rvert| \leq \kappa$,  and 
ii) by \eqref{desoer:upper:bound2}, $\lvert|Q(k)\rvert| \leq \frac{m_1^2}{1-p_1^2}$,  
it is clear from \eqref{sub:multi} that $\lvert|R_1\rvert| \leq 2\kappa\frac{m_1^2}{1-p_1^2}L$. Notice that \eqref{eq:diff:3} is a discrete-time Lyapunov equation; the solution for which is given by 
\begin{equation} \label{soln:lyapunov:1}
    Q(k+1) -Q(k)  = R_1 + \textstyle\sum_{j=1}^{\infty}[M_f^r(k)^\top]^j R_1 [M_f^r(k)]^j . 
    \end{equation} 
    \normalsize
  Taking the norm of both sides of \eqref{soln:lyapunov:1} leads to 
\begin{align}\label{eq:soln:lyapunov:1:norm:ineq:gp}
\lvert|Q(k+1) -Q(k)\rvert| & \leq \lvert|R_1\rvert|(1+ \textstyle\sum_{j=1}^{\infty}m_1^2p_1^{2j})\\
 & \leq 2\kappa\frac{m_1^4}{(1-p_1^2)^2}L.\label{eq:soln:lyapunov:1:norm:ineq}
\end{align} 
\normalsize
where inequality~\eqref{eq:soln:lyapunov:1:norm:ineq} is a consequence of \eqref{eq:soln:lyapunov:1:norm:ineq:gp} being a convergent series.  Next, pick $\sigma > 0$ such that \mbox{$1-\sigma <1$}. Hence, from inequality~\eqref{eq:soln:lyapunov:1:norm:ineq} it is clear that if $\kappa \leq \frac{(1-p_1^2)^2}{2m_1^{4}L}(1-\sigma)$, then $\lvert|Q(k+1) -Q(k)\rvert|  \leq 1-\sigma$. It turns out that  $\lvert|Q(k+1) -Q(k)\rvert|  \leq 1-\sigma$ implies, for $z^r \neq 0$ and 
$k \in \mathbb{Z}_{\geq 0}$, 
\begin{equation}\label{eq:show:neg:def}
z^r(k)^\top M_f^r(k)^\top Q(k+1)M_f^r(k) -Q(k)z^r(k)<0.
\end{equation} 
\normalsize
Indeed, note that \eqref{eq:Lyapunov} can be rewritten as:
   $ M_f^r(k)^\top Q(k+1)M_f^r(k) -Q(k) = -I_{n+q} + Q(k+1) -Q(k)$, for all $k \in \mathbb{Z}_{\geq 0}$. 
Therefore,  for all $k \in \mathbb{Z}_{\geq 0}$,
\eqref{eq:show:neg:def}, can be written as:
\begin{align} 
&z^r(k)^\top(-I_{n+q} + Q(k+1) -Q(k)) z^r(k) \nonumber \\
& \leq -\lvert|z^r(k)\rvert|^2 + z^r(k)^\top(Q(k+1) -Q(k))z^r(k) \nonumber \\
& \leq -\lvert|z^r(k)\rvert|^2 +
\lambda_{\max}(Q(k+1) - Q(k))\lvert|z^r(k)\rvert|^2  \label{neg:def:2:1} \\
& \leq -\lvert|z^r(k)\rvert|^2 + (1-\sigma)\lvert|z^r(k)\rvert|^2 \label{neg:def:3} \\
& = -\sigma \lvert|z^r(k)\rvert|^2  < 0, \label{neg:def:4}
\end{align}
\normalsize
where 
\eqref{neg:def:2:1} follows from the definition of the induced  norm of $(Q(k+1)-Q(k))^{\frac{1}{2}}$, 
\eqref{neg:def:3} is due to the following reasons: a) the norm of a matrix is lower bounded by its spectral radius \citep[Theorem~5.6.9]{horn2012matrix}, and b) $\lvert|Q(k+1) -Q(k)\rvert|  \leq 1-\sigma$,  and finally 
 \eqref{neg:def:4} follows from the assumption that $\sigma >0$. \\
 Therefore, by plugging~\eqref{eq:show:neg:def} in~\eqref{eq:lyap:diff:TV}, it is immediate  that   \scriptsize
\begin{align}
    & \Delta V(k,z^r)    
    <-2h(z^r)^\top M_f^r(k)^\top Q(k+1)
    \textstyle\sum_{\ell=1}^m X(z^\ell)B_f^rz^r \nonumber \\
    & ~~~+h^2(z^r)^\top B_f^r(k)^\top\textstyle\sum_{\ell=1}^m X(z^\ell) Q(k+1) \textstyle\sum_{\ell=1}^m X(z^\ell)B_f^r 
    z^r  \nonumber \\
    &=(z^r)^\top \big( h^2 B_f^r(k)^\top \textstyle\sum_{\ell=1}^{m}Z^\ell Q(k+1) \textstyle\sum_{\ell=1}^{m}Z^\ell B_f^r(k) \nonumber \\
 &~~~-2h^2B_f^r(k)^\top Q(k+1)\textstyle\sum_{\ell=1}^{m}Z^\ell B_f^r(k) \nonumber \\
 &~~~-2h (I-hD_f^r(k)) Q(k+1)\textstyle\sum_{\ell=1}^{m}Z^\ell B_f^r(k)\big)z^r \nonumber \\
 &\leq (z^r)^\top\big(h^2 B_f^r(k)\textstyle\sum_{\ell=1}^{m}Z^\ell Q(k+1) \textstyle\sum_{\ell=1}^{m}Z^\ell B_f^r(k) \nonumber \\ 
 &~~~~~-2h^2B_f^r(k)^\top Q(k+1)\textstyle\sum_{\ell=1}^{m}Z^\ell B_f^r(k) \big)z^r  \label{ineq:lyap:TV:hetero:1} \\
 & \leq (z^r)^\top\big(h^2 B_f^r(k)\textstyle\sum_{\ell=1}^{m}Z^\ell Q(k+1) \textstyle\sum_{\ell=1}^{m}Z^\ell B_f^r(k) \nonumber \\ &~~~-h^2B_f^r(k)^\top Q(k+1)\textstyle\sum_{\ell=1}^{m}Z^\ell B_f^r(k) \big)z^r  \label{ineq:lyap:TV:hetero:2} \\
 & {=}{-}(z^r)^\top\big(h^2B_f^r(k)^\top 
 (I{-}\textstyle\sum_{\ell=1}^{m}Z^\ell)Q(k{+}1)\textstyle\sum_{\ell{=}1}^{m}Z^\ell B_f^r(k)\big)z^r {\leq} 0,\label{ineq:lyap:TV:hetero:3}
\end{align}
\normalsize 
where inequality~\eqref{ineq:lyap:TV:hetero:1},~\eqref{ineq:lyap:TV:hetero:2}, and ~\eqref{ineq:lyap:TV:hetero:3} are obtained using the same line of reasoning as in inequality~\eqref{ineq:lyap:TV:1},~\eqref{ineq:lyap:TV:2}, and~\eqref{ineq:lyap:TV:3}, respectively. Exponential eradication of virus~$r$ with a domain of attraction $\mathcal D^r$ is a direct consequence of \citep[Theorem~28, Section~5.9]{vidyasagar2002nonlinear}. \qed 

\section{Numerical Analysis}

\begin{figure}[t]
  \centering
      \includegraphics[width=0.8\linewidth]{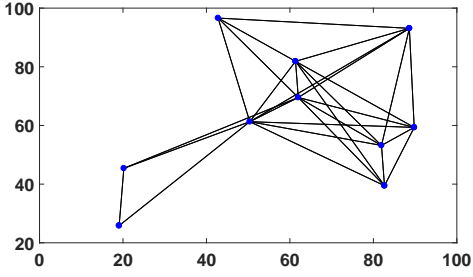}
      \caption{Graph structure for a network of individuals}
      \label{fig1}
\end{figure}

\seb{We consider a $10$-node network of individuals (i.e. $n=10$) on the network shown in Fig~\ref{fig1} with the edges having weights $a_{ij}$ equal to one. 
We consider a 5-node network of resources (i.e., $q=5$), with the network of resources being fully connected and the weights $\alpha_{ij}$, for all $i, j \in [5]$, is set to one.
Each node in the network of individuals is connected with each of the five resources, that is, $\beta_{ij}^{wr}=1$ for all $i \in [10]$, 
$j \in [5]$. 
Moreover, $c_{jl}^{wr}=1$  for every pair of $(j,l)$, where $j$ corresponds to the $j$th node in the resource network, and $l$ corresponds to the $l$th node in the population network.
We set $m=2$, i.e., two competing viruses.}

\seb{\textbf{Setting initial states:} For virus~$1$, $x_i^1(0) \in [0,0.5]$, for $i \in [10]$; $w_j^1(0) \in [0,2]$, for $j \in [5]$. For virus~$2$, $x_i^2(0) \in [0,0.4]$, for $i \in [10]$; and $w_i^1(0) \in [0,2]$, for $j \in [5]$. Choose the sampling period $h=0.001$. For all simulations, we plot the average infection level for a given virus in the network of individuals and that of resources.}


\seb{\textbf{Simulation for Theorem~1:} Choose, for $i \in [10]$, $\beta_i^1 = 0.01, \beta_i^{w1} = 0.01, \delta_i^1 = 3$,  and, for $j \in [5]$, $\delta_j^{w1} = 2$. Choose, for $i \in [10]$, $\beta_i^2 = 0.3, \beta_i^{w2} = 0.01, \delta_i^2 = 2$, and,  for $j \in [5]$, $\delta_j^{w2} = 1$. Observe that Assumptions 1--4 hold and $\rho(M_f^1)= 0.9984 < 1$, and $\rho(M_f^2)= 1.0012 > 1$. Figure~2 shows that consistent with Theorem~\ref{thm:global:DFE:TI}, the average infection level for virus~1 in the network of individuals and that of resources converge to zero; see the blue line and
green line, respectively. As an aside, notice that consistent with \citep[Theorem~3]{cui2022discrete}, the dynamics of virus~2 converge to an endemic equilibrium.   }


\begin{figure}[t]
  \centering
      \includegraphics[width=0.9\linewidth]{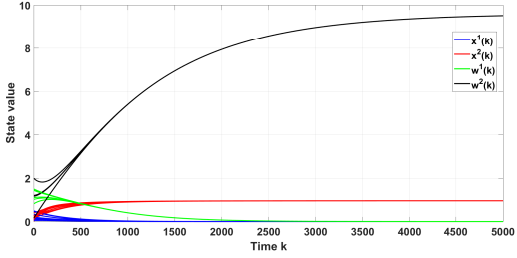}
      \caption{Time-invariant spread dynamics: eradication of virus~1; persistence of virus~2.}
      \label{fig2}
\end{figure}

\seb{\textbf{Simulation for Corollary~2:} Choose, for $i \in [10]$, $\beta_i^1 = 0.01, \beta_i^{w1} = 0.01, \delta_i^1 = 3$, and, for $j \in [5]$,  $\delta_j^{w1} = 2$. Choose, for $i \in [10]$, $\beta_i^2 = 0.005, \beta_i^{w2} = 0.01, \delta_i^2 = 2$, and, for $j \in [5]$, $\delta_i^{w2} = 1$.  Observe that Assumptions 1--4 hold, and $\rho(M_f^1)= 0.9984 < 1$, $\rho(M_f^2)= 0.9994 < 1$. Figure~3 shows that, consistent with Corollary~\ref{thm:DFE:exp:stable}, the average infection level for viruses~1 and~2 in the network of individuals and that of resources converge to zero.}


\begin{figure}[t]
  \centering
      \includegraphics[width=0.9\linewidth]{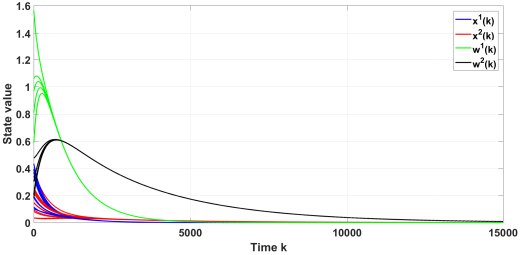}
      \caption{Time-invariant spread dynamics: both viruses get eradicated.}
      \label{fig4}
\end{figure}

\seb{\textbf{Simulation for Theorem~2:} Consider the following sets of values for the system parameters: 1)  for $i \in [10]$, $\beta_i^1(k) = 0.01, \beta_i^{w1}(k) = 0.01, \delta_i^1(k) = 3$, and, for $j \in [5]$, $\delta_j^{w1}(k) = 2$. For $i \in [10]$, $\beta_i^2(k) = 0.4, \beta_i^{w2}(k) = 0.01, \delta_i^2(k) = 2$, and, for $j \in [5]$, $\delta_j^{w2}(k) = 1$. 2) For $i \in [10]$, $\delta_i^1(k) = 0.01, \beta_i^{w1}(k) = 0.01, \delta_i^1(k) = 3$.  for $j \in [5]$, $\delta_j^{w1}(k) = 2$. For $i \in [10]$, $\beta_i^2(k) = 0.01, \beta_i^{w2}(k) = 0.01, \delta_i^2(k) = 2$, and, for $j \in [5]$, $\delta_j^{w2}(k) = 1$. For odd time instants, choose 1) for the parameter values; otherwise, choose 2). Assumptions 1, 6-8 hold and $\sup_{k \in \mathbb{Z}_{\geq 0}} \rho(M_f^1(k))= 0.9984 < 1$, $\sup_{k \in \mathbb{Z}_{\geq 0}} \rho(M_f^2(k))= 1.0023 > 1$. Figure~4 shows that the average infection level for virus~1 in the network of individuals and that of resources converge to zero (Theorem~\ref{thm:homogenous:spread}); see the blue line and the green line, respectively. It seems that if the condition is violated, then there exists an endemic equilibrium, to which the infection levels in the network of individuals and the shared resources converge; see the red and black lines, respectively.}

\begin{figure}[t]
  \centering
      \includegraphics[width=0.9\linewidth]{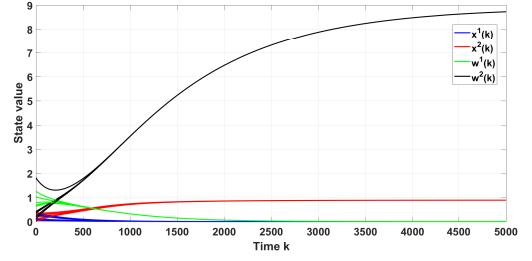}
      \caption{Time-varying spread dynamics with identical healing and infection rates for all individuals: virus~1 gets eradicated, but virus~2 remains persistent. }
      \label{fig6}
\end{figure}

\seb{\textbf{Simulation for Theorem~3:} The network of individuals is partitioned into two groups: Group $a$ (Node 1 -- Node 5) and Group $b$ (Node 6 -- Node 10). For nodes in group $a$ consider the following choices of parameter values: a1) for $i=1,2,\hdots 5$, $\beta_i^1(k) = 0.1, \beta_i^{w1}(k) = 0.01, \delta_i^1(k) = 3$,  and for $j \in [5]$, $\delta_j^{w1}(k) = 2$. For $i=6,7,\hdots 10$, $\beta_i^2(k) = 0.4, \beta_i^{w2}(k) = 0.01, \delta_i^2(k) = 2$, and for $j \in [5]$ $\delta_j^{w2}(k) = 1$. a2) For $i=1,2,\hdots 5$, $\beta_i^1(k) = 0.05, \beta_i^{w1}(k) = 0.01, \delta_i^1(k) = 3$, and for $kj \in [5]$, $\delta_j^{w1}(k) = 2$.  For $i=6,7,\hdots 10$ $\beta_i^2(k) = 0.2, \beta_i^{w2}(k) = 0.01, \delta_i^2(k) = 2$, and, for $j \in [5]$,  $\delta_j^{w2}(k) = 1$. For odd time instants; choose a1); otherwise, choose a2). For nodes in the group $b$, consider the following choices of values for the parameters: b1) For $i=1,2,\hdots, 5$, $\beta_i^1(k) = 0.1, \beta_i^{w1}(k) = 0.01, \delta_i^1(k) = 3$, and, for $j \in [5]$, $\delta_j^{w1}(k) = 2$. For $i=1,2,\hdots, 5$, $\beta_i^2(k) = 0.01, \beta_i^{w2}(k) = 0.01, \delta_i^2(k) = 2$, and, for $j \in [5]$, $\delta_j^{w2}(k) = 1$; b2) For $i=6,7,\hdots, 10$, $\beta_i^1(k) = 0.05, \beta_i^{w1}(k) = 0.01, \delta_i^1(k) = 3$, and, for $j \in [5]$, $\delta_j^{w1}(k) = 2$. For $i=6,7,\hdots, 10$, $\beta_i^2(k) = 0.01, \beta_i^{w2}(k) = 0.01, \delta_i^2(k) = 2$, and, for $j \in [5]$, $\delta_j^{w2}(k) = 1$.  For odd time instants, choose b1); otherwise, choose b2). Assumptions 6-8 hold and $\sup_{k \in \mathbb{Z}_{\geq 0}} \rho(M_f^1(k))= 0.9987 < 1$, $\sup_{k \in \mathbb{Z}_{\geq 0}} \rho(M_f^2(k))= 1.0023 > 1$. Further, $\lvert|M_f^1(k)\rvert| \leq 1.0016$ for all $k \in \mathbb{Z}_{\geq 0}$, and $\sup_{k \in \mathbb{Z}_{\geq 0}} \lvert|M_f^1(k+1)-M_f^1(k)\rvert| = 0.0005$. Figure~5 shows that, consistent with 
Theorem~\ref{thm:Main:Result2}, virus~1 is eradicated; see the blue and green lines, respectively. }

\begin{figure}[t]
  \centering
      \includegraphics[width=0.9\linewidth]{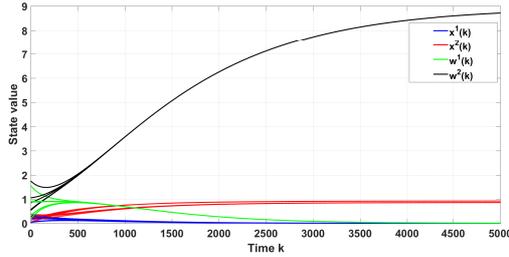}
      \caption{Time-varying spread dynamics with different healing and infection rates for some individuals: virus~1 gets eradicated, but virus~2 remains persistent.}
      \label{fig8}
\end{figure}

\section{Conclusion}

The paper studied the spread of multiple competing using a discrete-time time-varying multi-competitive layered networked SIWS model. For time-invariant graphs, we identified a sufficient condition for the exponential eradication of a virus. Thereafter, we established a sufficient condition for exponential eradication of a virus for spread over time-varying undirected graphs with all nodes having identical infection (resp. healing) rates. Finally, for spread over slowly time-varying (un)directed graphs with the nodes not necessarily having identical infection (or healing) rates, we provided a sufficient condition for exponential eradication of a virus. Future work should study the endemic behaviors of the proposed model. Moreover, identifying sufficient (resp. necessary) conditions for estimating the infection level in the population, given knowledge of infection levels in (a part of) the infrastructure network, remains an open problem.
\bibliography{References-Sebin}
\end{document}